\documentclass[preprint,proceedings]{rmaa}

\usepackage{paralist}

\usepackage{psfrag,color,psfig}

\newcommand{\ltapprox}{\raisebox{-0.5ex}{$\,\stackrel{<}{\scriptstyle\sim}\,$}}
\newcommand{\gtapprox}{\raisebox{-0.5ex}{$\,\stackrel{>}{\scriptstyle\sim}\,$}}
\newcommand{\lya}{\ifmmode {\rm Ly}\alpha \else Ly$\alpha$\fi}
\def\msun{$M_\odot$}
\def\lya{Lyman-$\alpha$}

\def\msun{\ifmmode M_{\odot} \else M$_{\odot}$\fi}
\def\zsun{\ifmmode Z_{\odot} \else Z$_{\odot}$\fi}
\def\lsun{\ifmmode L_{\odot} \else L$_{\odot}$\fi}
  
\SetYear{2004}
\SetConfTitle{
II International Workshop on Science with the GTC}

\title{Observing z$>7$ sources with the GTC} 

\author{
  R. Pell\'o,\altaffilmark{1} 
  D. Schaerer,\altaffilmark{2,1}
  J. Richard,\altaffilmark{1}
  J. -F. Le Borgne,\altaffilmark{1}
  and J. -P. Kneib\altaffilmark{3,1}}

\altaffiltext{1}{Laboratoire d'Astrophysique de l'Observatoire
Midi-Pyr\'en\'ees, France.}
\altaffiltext{2}{Observatoire de Gen\`eve, Switzerland.}
\altaffiltext{3}{Caltech Astronomy, USA.} 

\shortauthor{Pell\'o et al.}
\shorttitle{Observing z$>7$ sources with the GTC}

\fulladdresses{
\item J. -P. Kneib: Caltech Astronomy, MC105-24, Pasadena, CA 91125,
USA (\email{kneib@caltech.edu}).
\item J. -F. Le Borgne, R. Pell\'o and J. Richard: 
Laboratoire d'Astrophysique de l'Observatoire
Midi-Pyr\'en\'ees, UMR5572, 14 Av. Edouard-B\'elin, F-31400 Toulouse,
France (\email{leborgne, roser, jrichard@ast.obs-mip.fr}).
\item D. Schaerer: Observatoire de Gen\`eve,
51, Ch. des Maillettes, CH-1290 Sauverny, Switzerland
(\email{daniel.schaerer@obs.unige.ch}).
}

\listofauthors{R. Pell\'o, D. Schaerer, J. Richard, J. -F. Le Borgne \& J. -P. Kneib}
\indexauthor{Pell\'o, R.}
\indexauthor{Schaerer, D.}
\indexauthor{Richard, J.}
\indexauthor{Le Borgne, J. -F}
\indexauthor{Kneib, J. -P}

\abstract{We present the first results obtained from our pilot
ultra-deep near-IR survey with ISAAC/VLT, aimed at
the detection of z$>7$ sources using lensing clusters as natural 
Gravitational Telescopes. Evolutionary synthesis models of PopIII and
extremely metal-poor starbursts have been used to derive 
observational properties expected for these high-$z$
galaxies, such as expected magnitudes and colors, line
fluxes for the main emission lines, etc. These models have
allowed us to define fairly robust selection criteria to
find z$\sim 7-10$ galaxies based on broad-band near-IR
photometry in combination with the traditional Lyman
drop-out technique. The magnification in the core of lensing
clusters improves the search efficiency and subsequent 
spectroscopic follow up. The research efficiency 
will be significantly improved by
the future near-IR multi-object facilities such as EMIR/GTC and 
KMOS/VLT-2, and later the JWST.
}

\addkeyword{Cosmology: Early Universe}
\addkeyword{Galaxies: High Redshift}
\addkeyword{Galaxies: Evolution}
\addkeyword{Infrared: Galaxies}

\begin{document}
\maketitle

\section{Motivation}
\label{intro}

  The detection of the very first stars and galaxies forming from
pristine matter in the early Universe remains one of the major
challenges of observational cosmology (cf.\ review of Loeb \& Barkana
2001; Weiss et al. 2000; Umemura \& Susa 2001). 
These sources constitute the building blocks of galaxies,
forming at redshifts $z \sim$ 6--30. Indeed,
the recent WMAP results seem to place the very first generation of stars at
$z \sim$ 10--30 (Kogut et al. 2003). Important progress during the
last years has permitted direct observations of galaxies and quasars
out to redshifts $z \sim 6.6$  (Hu et al.\ 2002, Fan et al.\ 2003, 
Kodaira et al.\ 2003). At z$>7$, the most
relevant signatures are expected in the near-IR $\lambda \gtapprox 1
\mu$m window. Modeling efforts during the last years have been
motivated by JWST, which should be able to observe these objects at
redshifts up to $z \sim$ 30. However, the delay of JWST and the
availability of well suited near-IR facilities in ground-based 8-10m
class telescopes hasten the development
of observational projects targetting $z \gtapprox 7$ sources.
The first studies on the physical properties of these sources
can be presently started with instruments such as ISAAC/VLT, and continued 
with the future multi-object spectrographs such as EMIR at GTC 
($\sim$ 2006), or 
KMOS for the second generation of VLT instruments ($\sim$ 2009).

   In this paper we present the first results from our ongoing pilot
project with ISAAC/VLT, aiming at looking for the first stars/galaxies
through lensing clusters used as Gravitational Telescopes (GTs). 
The plan of the paper is the following. 
We summarize in section~\ref{properties} the observational properties
expected for galaxies at 7 $\ltapprox$ z $\ltapprox$ 10. In
section~\ref{VLTProject} we present the main characteristics of our 
ISAAC/VLT project. The method used to
identify sources and the first results obtained are also
presented in this section, in particular the first spectroscopic 
confirmation of a z=10 candidate. The implications of these results 
for EMIR/GTC are briefly discussed in section~\ref{EMIR}.

\section{Properties of 7 $\ltapprox$ z $\ltapprox$ 10 galaxies}
\label{properties}

   We have used the evolutionary synthesis models by Schaerer 
(2002, 2003) for Population III and
extremely metal deficient stars to derive the expected magnitudes and colors
of galaxies at 7 $\ltapprox$ z $\ltapprox$ 10. The dependence of their
properties on the IMF and upper mass limits for star formation were
studied in previous papers (e.g. Schaerer \&
Pell\'o 2001; Pell\'o \& Schaerer 2002). We have also computed the 
expected S/N ratios for the main emission lines using telescope and
instrumental parameters in the near-IR corresponding to ISAAC/VLT, and
the future multi-object facilities EMIR/GTC and KMOS/VLT.

\begin{figure}[!t]
  \includegraphics[width=\columnwidth]{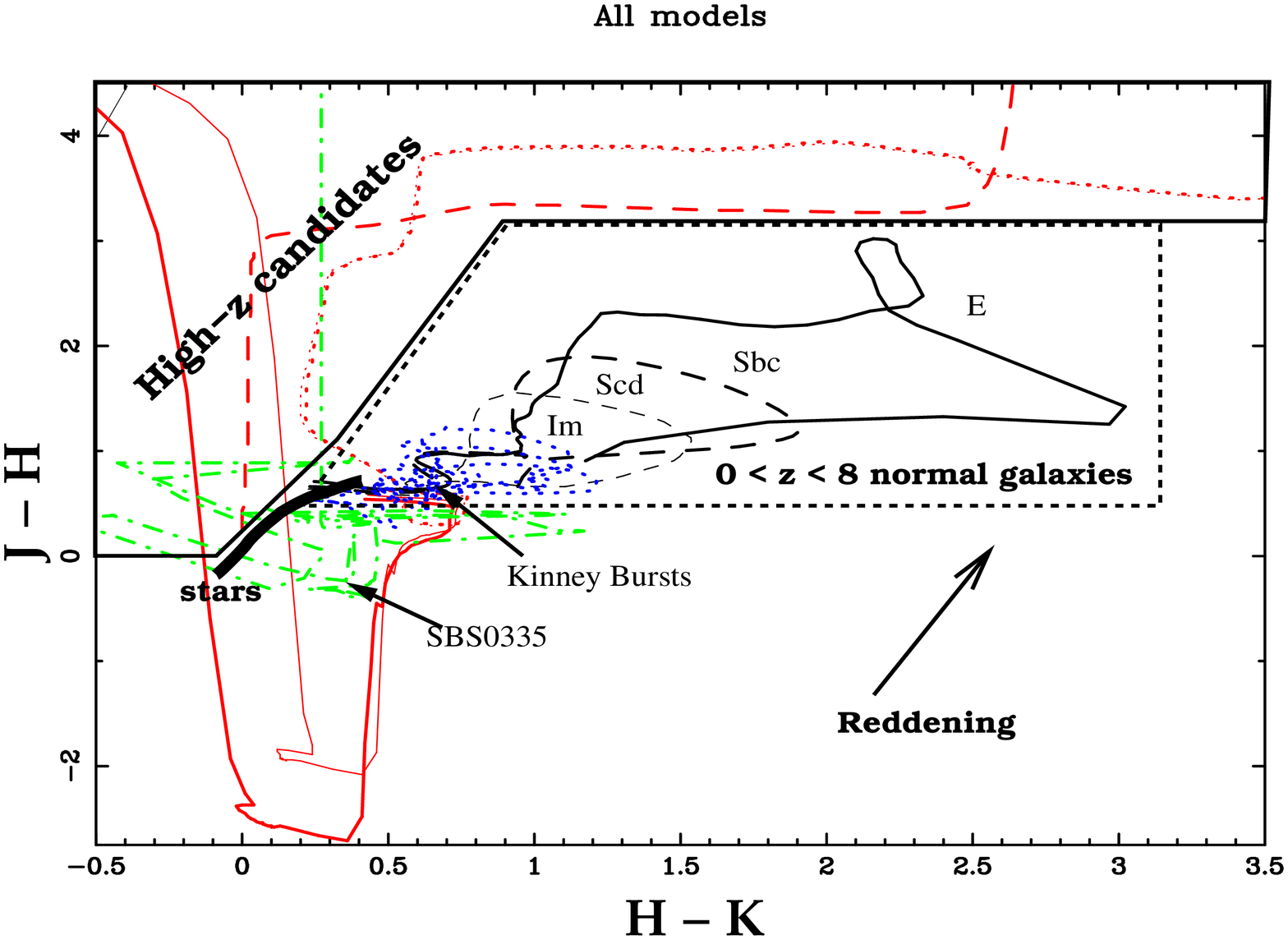}
  \caption{J-H versus H-Ks color-color diagram (Vega system)
for young starbursts (metal-poor and metal-free) objects, showing the
position for the different redshifts over the interval of $z \sim$ 5 to 12.
The position of stars and normal galaxies up to z $\le$ 8
including also starbursts templates (SB1 and SB2, from Kinney et al. 1996,
and the low metallicity galaxy SBS0335-052) are also shown,
as well as the shift direction induced by reddening.
Various line styles (solid, dashed, dotted) illustrate young starburst
models with \lya\ contributions of 100, 50, and 0 \%.
Because of the limiting magnitudes in the different filters, most
$z \sim$ 7--10  candidates are expected within the ``High-z
candidates'' region.
This diagram illustrates that such galaxies can well be separated
from ``normal'' objects on the basis of JHK photometry.
}
\label{fig_cc}
\end{figure}

   According to our simulations, the predicted magnitude in the 
K'(Vega) band is
typically $\sim$ 24 to 25 for the reference stellar halo mass of
$10^7$ M$_{\odot}$. Nebular continuous emission and strong emission
lines have important effects on the integrated fluxes and colors of
such galaxies. Figure~\ref{fig_cc} displays the J-H versus H-Ks
color-color diagram for different models with $5 \le z \le 11$,
compared to the location of stars and normal galaxies at $0 \le z \le
8$, including the starbursts templates SB1 and SB2, from Kinney et
al. 1996, and the low metallicity galaxy SBS0335-052. 
Broad-band colors do not allow
to constrain the physical properties of these galaxies, such as the
IMF, star-formation mass range, ages, etc., but they are useful to identify
sources on ultra-deep photometric surveys (cf. below).

   The main emission lines expected are Lyman $\alpha$ and HeII
$\lambda$1640, the strongest HeII line. In principle, 
Lyman $\alpha$ can be detected 
on near-IR spectroscopic surveys with 8-10~m class telescopes, 
with a reasonable S/N over the redshift intervals z $\sim$ 7 to 18, 
with some gaps depending on the spectral resolution (OH subtraction), 
the atmospheric transmission, the intrinsic properties of
sources and the intergalactic medium (IGM) transmission. A joint detection with HeII
$\lambda$1640 should be possible for z $\sim$  5.5-7.0 (Lyman $\alpha$ 
in the optical domain), z $\sim$ 7-14 with both lines
in the near-IR. A measure of the hardness of the ionising flux,
which constrains the upper end of the IMF and the age of the
system, could be obtained through the detection of both HeII
$\lambda$1640 and Lyman $\alpha$. A rough measurement of the continuum
using the photometric spectral energy distribution (SED) provides
additional information on the stellar populations, extinction, etc. 

\section{A pilot project with ISAAC/VLT }
\label{VLTProject}
 
   Lensing clusters acting as GTs are
useful tools to investigate the properties of distant galaxies. 
Successful systematic searches for $z \gtapprox 5$ galaxies with GTs have 
appeared during the last three years (e.g. Ellis et al. 2001; Hu et
al. 2002; Kneib et al. 2004). 

   We were granted ISAAC/VLT time to
develop a project aimed at searching for $z \gtapprox$ 7 galaxies using
GTs. This prototype observational program is presently going on. The
method used and the first results obtained are summarized below. 

\subsection{Method}
\label{Method}

  Galaxy candidates at $z \gtapprox 7$ are selected from
ultra-deep $JHKs$ images in the core of gravitational lensing
clusters for which deep optical imaging is also available, including
HST data. We apply the Lyman drop-out technique to deep optical
images. The main spectral discontinuity at $z \ga 6$ shortward of 
Lyman $\alpha$ is due to the large neutral H column density in the
IGM. In addition, galaxies are selected according
to Fig.~\ref{fig_cc}: we require a fairly red $J-H$
color (due to the discontinuity trough the $J$ band), and 
a blue $H-Ks$ colour corresponding to an intrinsically blue UV 
restframe spectrum. The detection in at least
two bands longward of Lyman $\alpha$  and the combination with 
the above $H-Ks$ colour criterion allows us to avoid contamination 
by cool stars.

      Ultradeep JHKs imaging of 2
lensing clusters were obtained: A1835 ($z=0.25$) and AC114 ($z=0.31$), under
excellent seeing conditions ($\sim 0.4'' $ to $0.6''$).
A complete report and analysis of these observations will be given
elsewhere (Richard et al.\, in preparation). 
The best candidates are retained for
follow-up spectroscopy with ISAAC/VLT. 
The broad band SED of each source
is used to constrain the redshift using a modified version of the public
photometric redshift code {\em Hyperz} (Bolzonella et al.\ 2000),
including numerous empirical and theoretical spectral templates for 
both galaxies and AGNs, and keeping internal extinction as a free
parameter. The J band domain, where \lya\ should be located for
objects in this redshift interval, is systematically explored with
ISAAC according to the priorities set by the photometric redshift
probability distribution. 

\begin{figure}[!t]
\includegraphics[width=\columnwidth]{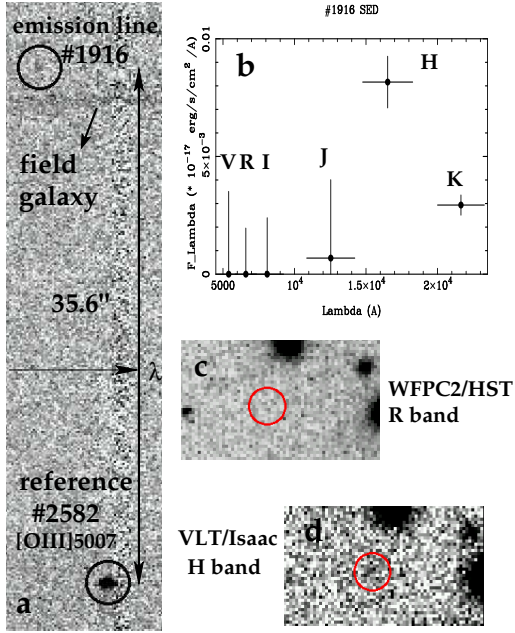}
\caption{The object $\#$1916 in A1835:
{\bf a)} 2D spectrum showing the detected emission line 
at 1.33745 \micron (leading to 
$z=10.00$ when identified
as \lya, as well the nearby field galaxy and the
[O~{\sc iii}] $\lambda$5007 line of the galaxy
\#2582 ($z=1.68$, Richard et al. 2003) used as a reference 
to stack the data sets. The line is seen on 2 independent 
overlapping bands. 
{\bf b)} Photometric optical to near-IR SED in $F_{\lambda}$ units. 
{\bf c) and d)} Thumbnail images in the HST-WFPC2 $R$ band, where the
object is not detected, and in the $H$ band.
}
\label{z10}
\end{figure}

\subsubsection{First Results on 7 $\ltapprox$ z $\ltapprox$ 10 
Photometric Candidates}

\begin{figure}[!t]
  \includegraphics[width=\columnwidth]{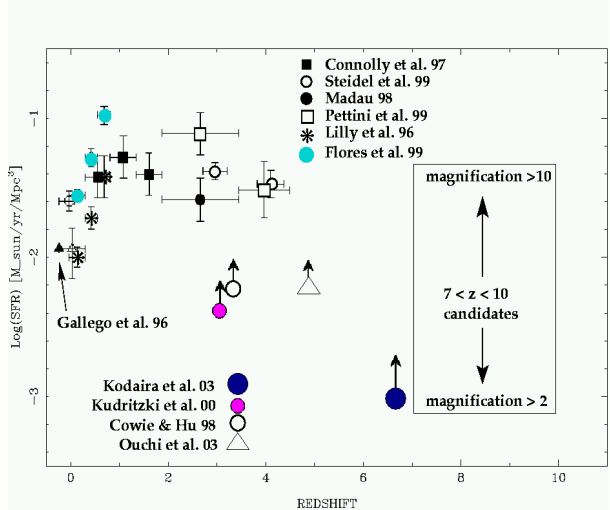}
  \caption{SFR density derived from our 
5 best candidates at 7 $\ltapprox$ z $\ltapprox$ 10 
in the lensing cluster A1835, compared to previous estimates up to 
$z \ltapprox 7$. Upper and lower limits correspond to extreme
assumptions for the typical magnification factors ranging between
$\ga$ 2 and $\ga$ 10.}
\label{SFR}
\end{figure}

Applying the above broad-band search criteria has provided 
5-10 $z > 7$ galaxy candidates per lensing 
cluster in the observed fields. About half of them have photometric
SEDs accurate enough to allow a robust identification within this redshift
interval, and thus to be included in the spectroscopic follow up sample.
Lensing effects are taken 
into account to determine the intrinsic properties of these sources
and the effective areas surveyed on the different source planes.
Typical magnification factors range between $\ga$ 2 and $\ga$ 10.
The corresponding number density of objects within 7 $\ltapprox$ z
$\ltapprox$ 10 ranges between $2 \times 10^{-2}$ and $4 \times$ 10$^{-4}$ per
Mpc$^{-3}$, i.e. typically a few 10$^{3}$ objects deg$^{-2}$. 
The main uncertainties are due to simplifying assumptions, 
incompleteness corrections, and the values adopted for the typical
magnification factors of our sample. 
From the observed magnitudes and magnification factors derived
from the lens model their star formation rates are estimated to be
a few \msun yr$^{-1}$ (lensing corrected, assuming a ``standard''
Salpeter IMF from 1--100 \msun).
Adopting typically $SFR \sim$ 2 \msun yr$^{-1}$ we obtain an estimate
of the star formation rate density shown in Fig.~\ref{SFR}, and compared to
previous estimates between $z=0-7$.
Obviously for now the SFR density at $z>7$ remains highly uncertain.
Ricotti et al. (2004) recently derived a fairly high SFR density
at $z=10$ based on our galaxy detection with spectroscopic redshift (A1835
IR1916, cf.\ below). However, their study suffers mostly from a highly
uncertain survey volume area and an overestimated {\em mean}
magnification factor. In short, a more detailed analysis and additional
observations are required to draw more firm conclusions
on the behaviour of the SFR density at $z>7$.

\subsubsection{Spectroscopic confirmation of a z=10 candidate}

We report in a recent paper the first likely spectroscopic
confirmation of a $z=10.0$ galaxy in our sample (see Pell\'o et
al. 2004). This galaxy (called \#1916) was the best candidate 
found in the field of the lensing cluster A1835, and a good example of
the research procedure. In this case, the photometric
redshift probability distribution shows a clear maximum at redshift
$z_{\rm phot} \sim$ 9--11. This estimate is mainly corroborated
by a strong break of $\ga$ 3.1 to 3.7 AB mags between
$VRI$ and $H$, which has a high significance independently of the definition
used for the limiting magnitudes (see Fig.~\ref{z10}).

   The spectroscopic redshift determination was obtained with ISAAC/VLT 
between 29 June and 3 July 2003, under excellent seeing conditions,
using a 1 arcsec slit width. The observed spectral interval covers the
range from 1.162 to 1.399 \micron (redshifts $z \sim$ 8.5--10.5 for
\lya). The observations resulted in the detection of one weak emission
line at the 4-5 $\sigma$ level with an integrated flux of 
$(4.1 \pm 0.5)\times 10^{-18}$ erg cm$^{-2}$ s$^{-1}$ at a 
wavelength of 1.33745 \micron\ (see Fig.\ \ref{z10}), which appears on
2 different overlapping wavelength settings. When 
identified as \lya\, in good agreement with the photometric SED, 
the observed line gives a redshift of $z=10.00175 \pm 0.00045$, the
most likely one given the data set (see a detailed discussion in
Pell\'o et 2004). 

   Given the location of the image on the lensing plane, the
source is highly magnified by a factor of 25. Its intrinsic AB
magnitude is 28.5 and 29 in the $H$ and $Ks$ bands respectively. 
The star formation rate obtained from a direct scaling of its 
UV restframe flux ranges between 1.8 and 3 \msun\ yr$^{-1}$ 
after lensing correction, assuming a ``standard'' Salpeter IMF 
from 1--100 \msun . The UV restframe spectrum is very blue, compatible
with a small extinction in this galaxy. Only a relatively small 
fraction of the \lya\ flux emitted from the source is observed.
Scaling young burst models with metallicities $Z \ga 1/50 \zsun$ 
to the observed SED we obtain $M_\star \sim  2 \times 10^8$ \msun\ 
(or lensing corrected values of $M_\star \sim 8 \times 10^6$ \msun) 
and luminosities of $L  \sim  4 \times 10^{11}$ \lsun\ 
($2 \times 10^{10}$ \lsun\ lensing corrected) for the standard
Salpeter IMF. The observed and derived properties of this object 
are in good agreement with expectations for a young protogalaxy 
experiencing a burst of star formation at $z=10$.

\section{Implications for EMIR/GTC}
\label{EMIR}

\begin{figure}[!t]
  \includegraphics[width=\columnwidth]{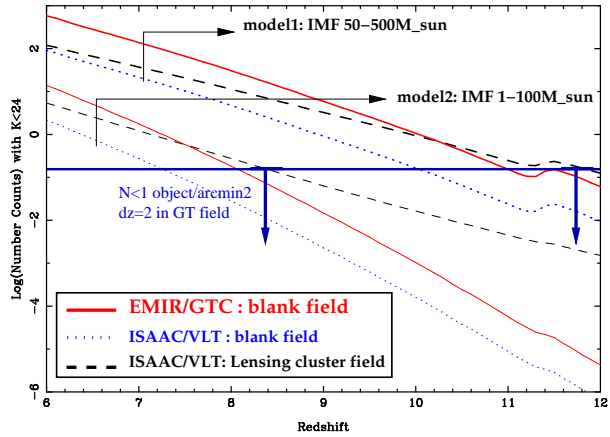}
  \caption{Comparison between the expected number counts of "primordial
galaxies" up to K$\le$24, per redshift $\Delta z=0.1$,
with and without the presence of a strong lensing cluster,
AC114 (z=0.31) in this example, 
within the field of view of ISAAC/VLT, compared to EMIR/GTC. 
Two different extreme assumptions for the IMF are
used. The solid horizontal line displays the domain corresponding to
less than 1 source per arcmin$^{-2}$ within $\Delta z =2$.
}
\label{counts}
\end{figure}

Simulations and pilot observations conducted by our group on the
possible detection of z $\gtapprox$ 7 galaxies support this scientific
case in view of the future ground-based near-IR facilities such as
EMIR at GTC and the future KMOS for the VLT 2nd generation (Schaerer \&
Pell\'o 2001; Pell\'o \& Schaerer 2002). An important issue of this 
project in view of the future facilities is the strategy for target
selection, and the efficiency achieved in spectroscopic studies. 
As shown in sections ~\ref{properties} and ~\ref{Method}, starbursts
could be detected from deep near-IR photometry based 
on a measurement of two colors with accuracies of the order of ~0.3-0.5
mag. As shown in Fig.~\ref{counts}, GTs are the ideal fields
for the first prospective studies.  We compare in this figure
the expected number counts of high-z galaxies with K'$<$ 24 in our GT
ongoing survey, for 2 different extreme assumptions for the IMF,
leading to intrinsically bright or faint sources. A simple
Press-Schechter formalism for the abundance of halos
and standard $\Lambda$CDM cosmology were considered in these
order of magnitude simulations, as well as a conservative
fixed fraction of the baryonic mass in halos converted into stars.

   As seen in Fig.~\ref{counts}, strong lensing fields are a factor of 
$\sim 5-10$ more efficient than blank fields of the same size in the z
$\sim 8-10 $ domain. This is the motivation of 
all present searches for z$\gtapprox$6-7 galaxies using
GTs (and ISAAC in particular). The large field of view of the future
instruments (between $\sim 6'-8'$ for EMIR/GTC and KMOS/VLT)
provides the same efficiency than GT fields in "optimistic" models,
and even better efficiency up to z$\sim$10, simply
because the size of the field compensates the magnification (and
dilution) effects. In the most pessimistic models (assuming
intrinsically fainter sources), the benefit of the wide field of view 
versus strong lensing extends only up to
z$\sim$7.5, the magnification being a major benefit in this case.
As shown in the figure, when this regime is attained, the number
of sources expected per redshift bin $\Delta z =2$ falls below 1
source per arcmin$^{-2}$, and thus we can still take benefit from 
the presence of GTs in the surveyed field. The redshift bin considered
($\Delta z =2$) corresponds to the typical wavelength interval which 
can be explored in a single shot with EMIR. 
These numbers are to be considered as orders of magnitude. Using
higher redshift GTs could improve the situation for "compact" lensing
studies with respect to $\sim 6'-8'$ blank fields. 

   Thanks to its multiplexing capabilities, spectral resolution and wide
field of view, EMIR at GTC will be the best suited instrument to
continue exploring the formation epoch of the first stars in galaxies
using significant samples of high$-$z sources.
A relatively high spectral resolution (R $\sim$ 3000-5000) 
considerably increases the
chances of detection between the sky lines. Once this is
obtained, the data can be rebinned to increase the S/N.
The medium spectral resolution is also favoured to attempt to measure the
emission line profiles, in particular to distinguish genuine
``primordial'' stellar sources from potential very high-z AGN
(cf. Tumlinson et al. 2001).  

\acknowledgements

We are grateful to A. Ferrara, M. Lemoine-Busserolle, D. Valls-Gabaud, 
T. Contini and F. Courbin for useful
comments and discussions. 
We thank the ESO Director General for a generous allocation of
Director's Discretionary Time for ISAAC spectroscopy (DDT 271.A-5013).
Also based on observations collected at the European Southern
Observatory, Chile (70.A-0355), the NASA/ESA Hubble Space Telescope
operated by the Association of Universities for Research in Astronomy, Inc.,
and the Canada-France-Hawaii Telescope operated by the
National Research Council of Canada, the French CNRS
and the University of Hawaii.
Part of this work was supported by the CNRS
and the Swiss National Foundation.

\adjustfinalcols

\end{document}